# Phonon Frequency and Its Modification by Magnon-Phonon Coupling from All-Temperature Theory of Magnon

Sambhu N. Datta
Department of Chemistry, Indian Institute of Technology, Powai, Mumbai – 400 076, India
Email: sndatta@chem.iitb.ac.in

**Abstract**

The all-temperature magnon (ATM) theory [J. Phys. Condens. Matter 21, 336003/1-14, 2009] has been used to analyze the temperature dependence of magnetization as well as the internal energy components of a mono-domain ferromagnetic solid. The critical exponents have been in better agreement with experiment than their mean-field theory and critical phenomenon theory counterparts, and unlike in the latter theories, vary from one ferromagnet to another. Expressions have been derived for the thermally averaged spin-center force constants and their break-up in terms of the base-line related (solid) and exchange-cum-field mediated (magnetic) components. These components give rise to expressions for phonon frequencies and their modifications by magnon-phonon coupling. The derived expressions are suitable for a correct quantum chemical evaluation of the involved properties. A detailed numerical calculation using spin configurations at varying crystal geometries is hardly possible even today and beyond the scope of the present work. The focus here is on the correctness and explaining the trends of properties. It has been shown that the frequency modification by magnon-phonon interaction can be negative for certain phonon branches near the ferromagnetic transition temperature. Also, the ratio of frequency modification and phonon frequency is approximately proportional to the ratio of curvatures of the involved energy surfaces.

## 1. Introduction

Magnon and phonon are quantized propagating excitations of the ordered magnetic moment and lattice vibrations. They can couple together to form a hybrid quasiparticle. Interaction of the stable magnon and phonon is an important topic of condensed matter physics today. This topic is further enriched by spectral complexities, coexistence of bulk and surface modes and the influence of the external magnetic field on magnon properties.

On the experimental side, coherent coupling between ferromagnetic magnons and superconducting qubits has been observed.[1] Also, magnon-phonon coupling has been directly evidenced in yttrium iron garnet.[2] Berk et al.[3] have directly validated strongly coupled magnon–phonon dynamics in a single thin nickel nanomagnet, using the fact that the coupling with phonons can lower the switching energy of nano-elements for manipulation and detection of magnons in them. They have also demonstrated that the magnon–phonon interaction can be tuned into the strong coupling regime by controlling orientation of the applied magnetic field. Of late, magnon-phonon interaction in spintronics has been reviewed covering a wide range of phenomena from interaction of coherent magnons with surface acoustic waves to the formation of magnon supercurrents in thermal gradients.[4]

Several authors have contributed to the development of theory. The magnon-phonon interaction in the continuum limit was described in references 5-7 as the phenomenological magnetoelastic energy to lowest order in the deviation of magnetization and that of the lattice from equilibrium. The Hamiltonian describing the magneto-elastic anisotropy was obtained by Rückriegel et al.[8] Bauer et al.[9] have investigated the effects of strong magneto-elastic coupling on the transport properties of magnetic insulators. They considered Boltzmann transport of the magnon-polarons, found transport coefficients as well as the spin diffusion length, and predicted anomalous features in the field and temperature dependence of the spin Seebeck effect in the case of adequately different disorder scattering in the magnetic and elastic subsystems. Shen[10] has relied on a mean field approach in a tight-binding model with nearest-neighbor approximation to study the magnon spectrum in yttrium iron garnet at a finite temperature. He found that the spin reduction from thermal excitations of magnons is different at two sets of iron atoms, the temperature dependence of the spin wave gap agrees well with experiment, and only two magnon modes are relevant to the ferromagnetic resonance. An elaborate theoretical work has been done by Streib et al.[11] These authors considered the magnon-phonon interactions to be mediated by both anisotropy and exchange. The Hamiltoanian of anisotropy from ref. 9 was used to obtain the form of anisotropy-mediated magnon-phonon interaction.[11] The effect of the latter on the phonon spectrum is small though it does not conserve the magnon number and leads to magnon decay. Using the exchange-mediated magnon-phonon interaction, the corresponding quasiparticle and transport lifetimes in magnetic insulators as well as the magnon field dependence have been determined. Streib et al.[11] found generally weak effect of phonon scattering on magnon transport and damping of Kittel's macrospin mode.[7-8]

Körmann et al.[12] computationally studied the temperature-dependent magnon-coupled phonon frequencies in bcc Fe. The authors referred to the spin-space averaged (SSA) density functional theory (DFT), [13] and adopted a drastic approximation for the temperature dependence of wave function. Explicit DFT calculations were done at the ferromagnetic ($T=0$) and paramagnetic ($T=\infty$) limits. For the intermediate region, the involved forces were determined from quantum Monte Carlo simulations for an effective nearest-neighbor Heisenberg model and subsequently used to define the parameter of temperature dependence. The phonon frequencies were estimated in the quasi-harmonic model, and the estimated magnon-phonon coupled frequencies were compared with experiment.

Below Curie temperature ($T_C$), magnetization is spontaneous and the material is ferromagnetic. Above Curie temperature $T_C$, the spontaneous magnetization vanishes, whereas the application of an external magnetic field causes a finite magnetization at $T_C$ that slowly decays as $T$ increases. Paramagnetism may be considered to firmly set in around the Curie-Weiss point $T_W$. The main point is that the ferromagnetic (FM) magnetization is driven by the interaction between neighboring spin centers as well as the interaction of the individual spin centers with the externally applied field $B_{app}$ and the molecular field $B_{mol}$, while the paramagnetic (PM) magnetization arises only from the interaction of the individual spin centers with $B_{app}$ and $B_{mol}$. Based on our previous formulation of an all-temperature magnon (ATM) treatment,[14] a theory of magnetization dependent on both an external field and temperature beyond the Curie point ($T > T_C$) has been prepared here. The temperature-dependent ATM wave function and its FM and PM counterparts have been defined and the corresponding magnetic energy expressions have been derived. A new expression for the thermal force constant has been formulated which is expected to generate a more realistic simulation of the magnon-phonon coupling.

The present article has been arranged as follows. The exchange Hamiltonian, the basic magnon theory and its shortcomings, and the correct form of the spin-wave Hamiltonian for use have been discussed in Section 2. Magnetic properties such as magnon population and magnetization have been formulated in Section 3 where an expression for the Curie-Weiss temperature $T_W$ for paramagnetism has been obtained also. Well-known characteristics of a mono-domain magnet such as the variation of magnetization as $T$ approaches $T_C$ from below and from above, and the dependence of susceptibility $\chi$ on temperature for $T > T_C$ have been discussed in Appendix 1. The critical exponents are to be in better agreement with experiment than their mean-field theory and critical phenomenon theory counterparts, and most important, they must vary from one ferromagnet to another. In Section 4, the ATM has been correlated with FM and PM pictures in different temperature regions, magnetization has been analyzed as a $T$-dependent linear combination of FM and PM magnetizations, and the ATM energy has been analyzed in terms of the FM energy, the PM energy and magnetization. In Section 5, the ATM force has been related to the $T$-dependent FM and PM forces, and the ATM force constant has been formulated as the sum of (base line) force constant for the "spin-less" solid and the (magnetic) force constant due to field and exchange. The phonon Hamiltonian has been written down and finally the phonon frequency and its change by magnon-phonon interaction have been formulated for correct evaluation. It is common knowledge that the phonon frequency decreases with lattice expansion as $T$ increases from absolute zero. It has been shown that the magnon modification of frequency can be negative for certain phonon branches at temperatures around $T_C$. The ratio of the two frequencies is approximately proportional to the ratio of the corresponding curvatures of energy surfaces in thermal equilibrium at any temperature. In Section 6, the achievements and limitations of the present work have been briefly discussed.

2. **Theoretical Background**

The traditional Heisenberg exchange operator $\mathbf{H}_{ex}(T)$ is the valence electron spin projected part of the total electronic Hamiltonian $\mathbf{H}_{el}(T)$ of a crystal of spin centers. The former operator is an effective spin Hamiltonian that conventionally describes the spin dynamics in a general ferromagnetic solid. It is written as[15-16]

$$\mathbf{H}_{ex} = -g\mu_B B \sum_{j=1}^{N} S_{jz} - \sum_{j,\varepsilon,\delta} J_\varepsilon \mathbf{S}_j \cdot \mathbf{S}_{j+\delta_\varepsilon} \tag{1}$$

where $g$ is the gyromagnetic ratio (taken as 2 for the electron spin), $B$ is the strength of the magnetic field, $\mu_B$ is the absolute magnitude of Bohr magneton given by $\mu_B = e_0\hbar/2m_ec$, $N$ is the number of spin centers per unit volume, and $J_\varepsilon$ is the nearest-neighbor coupling constant in the $\varepsilon$ direction. The indices $j$ and $\delta_\varepsilon$ represent the lattice point and the two unit vectors along positive and negative $\varepsilon$ directions respectively. The magnetic moment of each spin center with spin angular momentum $\hbar\mathbf{S}$ is given by $\boldsymbol{\mu} = g\mu_B\mathbf{S}$.

Calzado et al.[17-18] showed that the magnetic exchange integral $J$ consists of two contributions, $J = J_{FM} + J_{AFM}$. The ferromagnetic contribution is given by $J_{FM} = 2K$ where $K$ is the two-electron exchange integral, generally positive. The antiferromagnetic contribution is $J_{AFM} = [U - (U^2 + 16t_h^2)^{1/2}]/2$ that reduces to $-4t_h^2/U$ for $t_h \ll U$, where $t_h$ is the inter-site hopping energy and $U$ stands for the onsite effective two-electron interaction energy in the Hubbard model.

### A. Traditional approach

At this point one uses the Holstein-Primakoff transformation[19] that is a mapping from boson creation and annihilation operators to the spin operators and effectively truncates the infinite-dimensional Fock space of bosons to finite-dimensional spin subspaces:

$$S_j^+ = (2S)^{1/2} f(\hat{n}_j) a_j, \quad S_j^- = (2S)^{1/2} a_j^\dagger f(\hat{n}_j) \tag{2}$$

where

$$[a_j, a_{j'}^\dagger] = \delta_{j,j'}, \quad \hat{n}_j = a_j^\dagger a_j, \quad \text{and} \quad f(\hat{n}_j) = \left(1 - \frac{\hat{n}_j}{2S}\right)^{1/2}. \tag{3}$$

The boson operators $a_j$ and $a_j^\dagger$ are destruction and creation operators for magnons localized on site $j$ and $\hat{n}_j = S - \hat{S}_{jz}$ is the corresponding number operator showing that each magnon represents a spin flip from the microstate of the higher $m_s$ quantum number for each spin center. In terms of these new operators the exchange Hamiltonian in (1) can be written as

$$\mathbf{H}_{ex} = -g\mu_B B(NS - \hat{n}) - \left(\frac{NS}{2} - \hat{n}\right)\omega_0'$$
$$- S \sum_{j\varepsilon} J_\varepsilon \sum_{\boldsymbol{\delta}_\varepsilon} \left[ f(\hat{n}_j) a_j a_{j+\boldsymbol{\delta}_\varepsilon}^\dagger f(\hat{n}_{j+\boldsymbol{\delta}_\varepsilon}) + a_j^\dagger f(\hat{n}_j) f(\hat{n}_{j+\boldsymbol{\delta}_\varepsilon}) a_{j+\boldsymbol{\delta}_\varepsilon} \right] \tag{4}$$
$$- \sum_{j\varepsilon} J_\varepsilon \sum_{\boldsymbol{\delta}_\varepsilon} \hat{n}_j \hat{n}_{j+\boldsymbol{\delta}_\varepsilon}$$

where $\bar{J} = 2z^{-1} \sum_\varepsilon J_\varepsilon$, $\omega_0' = 2zS\bar{J}$ and $\hat{n} = \sum_j \hat{n}_j$. Similar use can be made later with the definitions $\bar{J}_{\boldsymbol{k}} = z^{-1} \sum_\varepsilon J_\varepsilon \sum_{\boldsymbol{\delta}_\varepsilon} e^{i\boldsymbol{k}\cdot\boldsymbol{\delta}_\varepsilon}$, $\omega_{\boldsymbol{k}}' = 2zS\bar{J}_{\boldsymbol{k}}$.

It is possible to show[14]

$$\langle \mathbf{H}_{ex} \rangle_n = -g\mu_B B(NS - n) + \frac{n(2NS - n)}{2NS}\omega_0' - \frac{1}{2NS} \sum_{\boldsymbol{k}\neq 0} |\sum_{\substack{j=1, n_j=0 \\ (n_1+\ldots+n_N=n)}}^{N, 2S} e^{-i\boldsymbol{k}\cdot\boldsymbol{R}_j}|^2 + \ldots \tag{5}$$

so that in zero field ($B=0$), $\langle \mathbf{H}_{ex} \rangle_{\{2S-n_j\}} = \langle \mathbf{H}_{ex} \rangle_{\{n_j\}}$, that is, there is no directional preference.

A propagating magnon is represented by the operators

$$b_{\boldsymbol{k}} = N^{-1/2} \sum_{j=1}^N e^{i\boldsymbol{k}\cdot\boldsymbol{R}_j} a_j,$$
$$b_{\boldsymbol{k}}^\dagger = N^{-1/2} \sum_{j=1}^N e^{-i\boldsymbol{k}\cdot\boldsymbol{R}_j} a_j^\dagger \tag{6}$$

such that $[b_{\boldsymbol{k}}, b_{\boldsymbol{k'}}^\dagger]_- = \delta_{\boldsymbol{k},\boldsymbol{k'}}$, $b_{\boldsymbol{k}}^\dagger b_{\boldsymbol{k}} = \hat{n}_{\boldsymbol{k}}$ and $\hat{n} = \sum_{\boldsymbol{k}} \hat{n}_{\boldsymbol{k}}$. The wave vector is $k_\varepsilon = 2\pi s/N_\varepsilon \delta_\varepsilon$ where $s$ varies in the range $-N_\varepsilon/2 + 1 \leq s \leq N_\varepsilon/2$. The site magnon operators are inverse Fourier transforms of the propagating magnon operators so that $\hat{n}_j = N^{-1} \sum_{k,l} e^{i(k-l)\cdot R_j} b_k^\dagger b_l$. If one substitutes the inverse transforms in equation (4), one obtains the exchange Hamiltonian for propagating magnons. The total number operator is

$$\hat{n} = \sum_j^{sites} \hat{n}_j = \sum_k^{\substack{magnon \\ modes}} \hat{n}_k \qquad (7)$$

and both $\mathbf{H}_{el}$ and $\mathbf{H}_{ex}$ commute with it.

Traditional applications to non-metallic molecular crystals can be seen from references 20-21.

### B. Traditional difficulties

Equation (4) reveals the nuances of the magnon formalism. First, it is customary to expand the square root operator $f(\hat{n}_j)$ into a series involving increasing powers of $(\hat{n}_j/2S)$, and the series rapidly loses convergence as temperature rises and $n_j$ increases past $S$. This limitation can be overcome by using an alternative spin representation in terms of non-hermitian Dyson-Maleév variants, though at the cost of the Hamiltonian having a complex appearance and the formalism losing its lucid interpretations.[22-23] Second, the boson space at each site is infinitely stretched so that unphysical states having population $n_j > 2S$ can mix in the treatment. Third, it is normal to apply a cut-off at the second order of magnon-magnon interaction, leaving out a major part of the series expansion. Because of these reasons, the magnon formalism is said to break down at the critical temperature $T_C$ where the thermally averaged magnon population at each site equals $S$. This must not generate the erroneous idea that at $T_C$ the coupling constant $J$ suddenly decays to zero. The latter depends on the crystal structure and the FM to PM transition is hardly accompanied by a drastic change in crystal structure.

An accurate evaluation of the spin-state statistics is not possible with the standard spin-wave formalism. Ref. 14 considers a spin-wave treatment of ferromagnetism that takes into account a finite yet exact series expansion of the square root operator, and each statistical sum is carried out only over the finite number of physical states so that the analysis becomes valid for all ranges of temperature.

### C. The Correct Spin-Wave Hamiltonian

Kubo[24] in his treatment of classical spin waves resorted to an exact finite series expansion of the square root function $f(n) = [1-n/2S]^{1/2}$ where $n = 0, 1, \ldots, 2S$: $f(n) = \sum_{i=0}^{2S} f_i n^i$. For different $S$, the coefficients $f_i$'s are precisely known, with $f_0 = 1$ for all $S$. Because $f(\hat{n}_j)$ operates on the state vectors in number space, one can use the unit projector in number space

$$\hat{\mathbf{1}} = \prod_{i=1}^{N} \sum_{n_i=0}^{2S} |n_i\rangle\langle n_i| \qquad (8)$$

to get the operator equality

$$f(\hat{n}_j) = \sum_{i=0}^{2S} f_i \hat{n}_j^i. \tag{9}$$

The correct spin-wave Hamiltonian operator is obtained when this equivalence is inserted into the Hamiltonian in (4),

$$\begin{aligned}\mathbf{H}_{ex} = &-g\mu_B B(NS - \hat{n}) - \left(\frac{NS}{2} - \hat{n}\right)\omega_0' \\ &- S\sum_{j\varepsilon} J_\varepsilon \sum_{\boldsymbol{\delta}_\varepsilon} \sum_{p,q=0}^{2S} f_p f_q \left[\hat{n}_j^p a_j a_{j+\boldsymbol{\delta}_\varepsilon}^\dagger \hat{n}_{j+\boldsymbol{\delta}_\varepsilon}^q + a_j^\dagger \hat{n}_j^p \hat{n}_{j+\boldsymbol{\delta}_\varepsilon}^q a_{j+\boldsymbol{\delta}_\varepsilon}\right] \\ &- \sum_{j\varepsilon} J_\varepsilon \sum_{\boldsymbol{\delta}_\varepsilon} \hat{n}_j \hat{n}_{j+\boldsymbol{\delta}_\varepsilon}.\end{aligned} \tag{10}$$

This operator would not generate any unphysical effect when only the physically relevant number states, that is, those with $0 \leq n_j \leq 2S$ for all $j$, are chosen as basis. Thus the magnon picture does not end at Curie point – the so-called breaking down is merely an artifact of the power series expansion and the choice of states – although the spontaneous magnetization vanishes at $T_C$. The inverse Fourier transforms may be substituted in equation (10) to get the correct exchange Hamiltonian for propagating magnons [equation (12) of reference 14]:

$$\mathbf{H}_{ex} = \mathbf{H}^0 + \mathbf{H}',$$

$$\mathbf{H}^0 = -g\mu_B BNSB + g\mu_B B\sum_{k}\hat{n}_k - zNS^2\bar{J} + \sum_{k}\hat{n}_k \omega_k,$$

$$\mathbf{H}' = -\frac{1}{2}\sum_{p,q=0}^{2S}\frac{f_p f_q}{N^{p+q}} \times$$

$$\sum_{\substack{k_1,\ldots,k_p\\l_1,\ldots,l_p}}\sum_{\substack{k_1',\ldots,k_q'\\l_1',\ldots,l_q'}}\sum_{m}\left[\omega'_{m-\sum(k-l)}b_{k_1}^\dagger b_{l_1}\ldots b_{k_p}^\dagger b_{l_p} b_m b_{m-\sum(k-l)-\sum(k'-l')}^\dagger b_{k_1'}^\dagger b_{l_1'}\ldots b_{k_q'}^\dagger b_{l_q'}\right. \tag{11}$$

$$\left.+\omega'_{m+\sum(k'-l')}b_m^\dagger b_{k_1}^\dagger b_{l_1}\ldots b_{k_p}^\dagger b_{l_p} b_{k_1'}^\dagger b_{l_1'}\ldots b_{k_q'}^\dagger b_{l_q'} b_{m+\sum(k-l)+\sum(k'-l')}\right]$$

$$-\sum_{k,l,m}\left[\frac{f_1}{N}(\omega_k' + \omega_l') + \frac{1}{2NS}\omega_{k-l}'\right]b_k^\dagger b_m^\dagger b_l b_{k-l+m}.$$

Both $\mathbf{H}^0$ and $\mathbf{H}'$, and therefore $\mathbf{H}_{ex}$, are dependent on temperature as the coupling constants rely on the lattice constants and crystal structure. The base line $E_{crystal}$ also depends upon temperature. It equals the difference between total (DFT) energy of the crystal at (approximately) the Born-Oppenheimer geometry in thermal equilibrium and the ATM magnetic energy from the exchange Hamiltonian, both at temperature $T$. Total energy under ATM may be estimated from Monte Carlo simulations of spin configurations, calculating the corresponding DFT energies and averaging with the appropriate thermal weights.

### 3. Magnetic properties

#### A. Magnon population

With the restriction on population of the localized states, one can consider the population in the individual propagating mode to be constrained as $0 \leq n_k \leq 2S$ (while neglecting the sum of off-diagonal terms in the expansion of $\langle \hat{n}_k \rangle$ in thermal equilibrium). The magnetization is written as

$$M = g\mu_B(NS - \langle \hat{n} \rangle_T) = (1 - \langle \hat{n} \rangle_T / NS)N\mu \tag{12}$$

where $\langle \hat{n} \rangle_T$ is the total number of magnons in thermal equilibrium at temperature $T$. The zero order operator $\mathbf{H}^0$ gives

$$\langle \hat{n} \rangle_T^{(0)} = NS[1 - \mathscr{B}_S(\theta)] \tag{13}$$

in terms of the Brillouin function

$$\mathscr{B}_S(\theta) = \frac{2S+1}{2S} ctnh\left(\frac{(2S+1)\theta}{2S}\right) - \frac{1}{2S} ctnh\left(\frac{\theta}{2S}\right) \tag{14}$$

where the variable is given by[14]

$$\theta = -S \log_e x^{(0)},$$
$$x^{(0)} = \frac{\phi^{(0)}(1)}{N} = \frac{3}{2}\left(\frac{3}{5}\right)^{3/2} e^{-g\mu_B B/\tau^{5/3}} \int_0^\infty e^{-\bar{\varepsilon}_1 y / 2S\tau} y^{1/2} dy. \tag{15}$$

Here, $\phi^{(0)}(1)$ is the distribution function for the one-magnon states in the grand canonical ensemble, $\tau = k_B T$, and $\bar{\varepsilon}_1$ is the average zeroth order one-magnon energy in the long wavelength limit, given by

$$\bar{\varepsilon}_1^{(0)} = \tilde{\Gamma}\omega_0',$$
$$\tilde{\Gamma} = \frac{2\pi^2}{5}\left(\frac{3}{4\pi}\right)^{2/3} \Gamma, \tag{16}$$

while the parameter $\Gamma$ is of order unity and its exact value depends on the Bravais lattice as shown in Table **S1** of Supplementary Information (SI). Because of thermal expansion, the coupling constants and $\omega_0'$ and hence $\bar{\varepsilon}_1^{(0)}$ generally decrease with increasing temperature. The main effect of the perturbation $\mathbf{H}'$ is to generate a temperature-dependent effective one-magnon energy $\bar{\varepsilon}_1$. The latter is less than $\bar{\varepsilon}_1^{(0)}$ because of increasing magnon-magnon and magnon-phonon scattering; anisotropy also contributes.

### B. Magnetization

If one combines equations (12) and (13), one finds the ratio of magnetization to its saturation value,

$$m = M/N\mu = \mathscr{B}_S(\theta) = \kappa\theta(1 - \eta\theta^2 + \sigma\theta^4 + ...) \tag{17}$$

where $\kappa = (S+1)/3S$, $\eta = [(2S+1)^2+1]/60S^2$, $\sigma = [(2S+1)^4+(2S+1)^2+1]/2520S^4$, etc. A list of constants $\kappa$, $\eta$ and $\sigma$ for different $S$ is given in Table **S2** of SI.

Expanding $x^{(0)}$ from (15) into a power series involving $a = \bar{\varepsilon}_1/2S\tau$ and after some more calculation one finds

$$\theta = Sa\left[1 - \frac{2}{21}a - \frac{8}{1701}a^2 - \frac{188}{130977}a^3 - ...\right] + \frac{\mu B}{\tau}. \tag{18}$$

As $T$ increases, $a$ decreases faster than $\bar{\varepsilon}_1$ so that near $T_C$ and above, $\theta \to Sa + \mu B/\tau$. According to the molecular field theory, the net field is a sum of the applied field $B_{app}$ and the molecular field $B_{mol}$ such that $B = B_{app} + B_{mol} = B_{app} + \lambda M$, $\lambda$ being the molecular field constant. Therefore,

$$\theta \to [\bar{\varepsilon}_1/2 + (\mu B_{app} + mN\mu^2\lambda)]/\tau. \tag{19}$$

In absence of an external field and as $t = T/T_C \to 1-$, $m \to 0$ and $\bar{\varepsilon}_1/\tau \ll 1$ so that $\theta \ll 1$, $m \sim \kappa\theta$ and $\tau_C = k_B T_C = N\mu^2\lambda\kappa$. Around $T_C$ and at higher temperatures, when the applied field is finite, $\bar{\varepsilon}_1/\tau$ is no longer negligibly small and

$$\theta \to \bar{\varepsilon}_1/2\tau + \mu B_{app}/\tau + m/t\kappa = (m+\xi)/\kappa t \tag{20}$$

where

$$\xi(T) = [\bar{\varepsilon}_1(T)/2 + \mu B_{app}]\kappa/\tau_C. \tag{21}$$

With equation (20), relation (17) can be written as

$$m(t-1) = \xi - \frac{\eta}{(\kappa t)^2}(m+\xi)^3 + \frac{\sigma}{(\kappa t)^4}(m+\xi)^5 + .... \tag{22}$$

Furthermore,

$$\xi_C = \xi(T = T_C) = [\bar{\varepsilon}_1(T_C)/2 + \mu B_{app}]\kappa/\tau_C,$$
$$\xi_P = \xi(T = \infty) = \mu B_{app}\kappa/\tau_C, \tag{23}$$
s.t. $\xi(T) = \mathcal{Q}(T)\xi_C + [1 - \mathcal{Q}(T)]\xi_P.$

All information of the magnon-phonon coupling are inherent in the decay of $\xi(T)$ from $\xi_C$ to $\xi_P$ as $T$ varies from $T_C$ to $\infty$. The decay of $\xi(T)$ stems from the decay of the magnetic exchange coupling constants that in turn leads to the decay of $\bar{\varepsilon}_1(T)$, and can be visualized as being exponential in nature: $\mathcal{Q}(t) = e^{-\zeta(t-1)}$.

### C. The Weiss temperature

To get an expression for the Weiss temperature, one can approximate equation (22) to write $m(T/T_C - 1) \approx \xi_C$ and $m(T/T_W - 1) \approx \xi_P$ for the same $m$ so that the estimated $t_P = T_W/T_C$ behaves as $t_P = [1 - (\xi_C - \xi_P)/(m + \xi(T))]^{-1}$. This would be truly valid in the case $T \to \infty$ where $m \to 0$ so that

$$t_P = [1 - (\xi_C - \xi_P)/\xi(\infty)]^{-1}. \tag{24}$$

The following results are expected from numerical calculations: (i) If $\xi(T)$ is fixed at $\xi = \xi_C$ in the above-$T_C$ range, then $t_P$ would be found as $t_P = \xi_C/\xi_P = 1/x_{PM}$ where $x_{PM}$ is the ratio $\xi_P/\xi_C$. (ii) Alternatively, if $\xi(T)$ is varied such that $\xi(\infty) = \xi_P$, then one gets $t_P = x_{PM}/(2x_{PM} - 1)$. The latter value is slightly larger than $1/x_{PM}$. The calculated plot of $1/\chi$ versus $T$ is included in Appendix 1.

From equation (17), one gets the temperature-dependence of $m$ for temperature approaching the Curie point from below ($t \to 1-$) in absence of an applied field. Relation (22) would help in studying the magnetic susceptibility in the region $T > T_C$ where magnetization occurs only in a finite applied field ($B_{app} \neq 0$). These properties are discussed in Appendix 1.

## 4. Correlation of ATM with FM and PM

### A. General ATM forms

Both $\mathbf{H}_{ex}$ and the total number operator $\hat{n}$ can have individual number states as common eigenstates. This is instrumental in finding the form of the wave function $\Psi_g$ for thermal ground state. In general, $\Psi_g$ and the corresponding total number of magnons and total magnetic energy at temperature $T$ (even beyond $T_C$ and for $B_{app} \neq 0$) are given by

$$\Psi_g(T) = \left[Tr(e^{-\mathbf{H}_{ex}(T)/\tau})\right]^{-1/2} \sum_{n_k,n_l,\ldots=0}^{2S} e^{-\in_{Magnon}(n_k,n_l,\ldots|T)/2\tau} |n_k, n_l, \ldots\rangle,$$

$$\langle \hat{n} \rangle_T = Tr\{e^{-\mathbf{H}_{ex}(T)/\tau}\hat{n}\} / Tr\{e^{-\mathbf{H}_{ex}(T)/\tau}\}, \tag{25}$$

$$\overline{\in}_{Magnon}(T) = Tr\{e^{-\mathbf{H}_{ex}(T)/\tau}\mathbf{H}_{ex}(T)\} / Tr\{e^{-\mathbf{H}_{ex}(T)/\tau}\}.$$

From equation (18) of ref. 14,

$$\in_{Magnon}(n_k,n_l,\ldots|T) = -g\mu_B B(NS-n) - zNS^2\overline{J} + \varepsilon_n^{(0)}(T). \tag{26}$$

The thermal average of magnon energy in (26) contains $n$ in place of $\langle \hat{n} \rangle_T$. In the long wavelength limit, $\varepsilon_n^{(0)}(T) = n\varepsilon_1^{(0)}(T)$ with $\varepsilon_1^{(0)}(T) \sim \overline{\varepsilon}_1^{(0)}(T)/2S \sim \overline{\varepsilon}_1(T)/2S$. Zeroth order expressions for the magnon number and $\overline{\in}_{Magnon}(T)$ in a non-vanishing applied field were derived in Ref. 14. See equation (34) for magnon number and equations (18), (23) and (24) for energy in the same reference. The zeroth-order expressions are good approximations when the magnon-magnon interaction is weak and its higher order effects can be neglected. Furthermore, the thermal ground state energy is given by

$$E_g(T) \doteq E_{crystal}(T) + \overline{\in}_{Magnon}(T). \tag{27}$$

Total energy $E_g$ and the base line $E_{crystal}$ may be calculated as follows. First, DFT total energies for selected spin configurations are computed under Born-Oppenheimer approximation for a set of crystal geometries. Then for each temperature, the thermal energy average is to be calculated for each specific geometry. The geometry is optimized for each specific $T$ and the optimized average energy is written as $E_g(T)$. Finally, the $\overline{\in}_{Magnon}(T)$ is calculated at the optimized geometry and $E_{crystal}(T)$ can be estimated.

### B. Temperature regions

It would be convenient to divide the whole temperature range into two distinct regimes, namely, the FM regime with spontaneous magnetization below the Curie point $T_C$, and the mixed regime with only a faint remembrance of the FM behavior at $T > T_C$ and mostly having a PM facet at temperatures greater than the Weiss temperature $T_W$.

Region I is $0 \leq T \leq T_C$ for which $B_{app}$ is chosen 0. This gives the wave function, magnon number and the internal energy of magnetism as

$$\Psi_g(T) \equiv \Psi_{FM}(T)$$
$$= \left( \sum_{n_k,n_l,...=0}^{2S} e^{-n(\bar{\varepsilon}_1/2S + g\mu_B B_{mol})/\tau} \right)^{-1/2} \sum_{n_k,n_l,...=0}^{2S} e^{-n(\bar{\varepsilon}_1/2S + g\mu_B B_{mol})/\tau} |n_k, n_l, ...\rangle,$$

$$\langle \hat{n} \rangle_T = \sum_{n_k,n_l,...=0}^{2S} n e^{-n(\bar{\varepsilon}_1/2S + g\mu_B B_{mol})/\tau} \bigg/ \sum_{n_k,n_l,...=0}^{2S} e^{-n(\bar{\varepsilon}_1/2S + g\mu_B B_{mol})/\tau}, \quad (28)$$

(with $n = n_k + n_l + ...$ and $B_{mol} = \lambda_{mol} M$)

$$\bar{\in}_{Magnon}(T) \equiv \in_{FM}(T) = N\left(-zS^2 \bar{J} + (\bar{\varepsilon}_1/2)(1-m) - (\tau_C/\kappa)m^2\right).$$

The first two terms in magnon free energy evolve from the interaction of the molecular field with magnetization. The second and fourth terms linearly vary with magnon number.

In range II, $T_C \leq T \leq \infty$, $B_{app} \neq 0$ and $\varepsilon_1/\tau \ll 1$ so that the exchange Hamiltonian can be written as

$$\mathbf{H}_{ex}(T) \doteq -(\mu BN + zNS^2 \bar{J}(T)) + \left(\frac{\bar{\varepsilon}_1(T)}{2} + \mu B\right)\frac{\hat{n}_T}{S}, \quad (29)$$

$B = B_{app} + m(T)\tau_C/\kappa\mu$.

Using the relation $\theta \cong (m+\xi)/t\kappa$, the thermal ground state wave function can be reformulated,

$$\Psi_g(T) = \prod_k |\psi_k(T)\rangle,$$

$$|\psi_k(T)\rangle = \left(\frac{1-x}{1-x^{2S+1}}\right)^{1/2} \sum_{n_k=0}^{2S} x^{n_k/2} |n_k\rangle \quad (30)$$

where $x = e^{-\theta(T)/S}$. From this wave function one obtains equation (13) analogue $\langle \hat{n} \rangle_T = NS[1 - \mathcal{B}_S(\theta)]$ that gives equation (17) for $m(T)$, and the following magnetic internal energy per unit volume

$$\bar{\in}_{Magnon}(T) \doteq N\left[-zS^2 \bar{J}(T) + \frac{1}{2}\bar{\varepsilon}_1(T)(1-m(T)) - \frac{\tau_C}{\kappa}m^2(T) - m(T)\mu B_{app}\right]. \quad (31)$$

Here, the fourth term in the magnon energy represents field effect. All other terms are to be interpreted as before.

### C. Magnetic analysis for $T > T_C$

Magnetization reveals an intertwined trend in (22). However, as $m \ll 1$ and $\xi(t)$ is also small for $\mu B_{app}/\tau \ll 1$, one can use $m \sim \xi/(t-1)$ and write

$$m(t) \doteq \mathcal{C}(t)m_C(t) + [1 - \mathcal{C}(t)]m_P(t) \quad (32)$$

and

$$m_C(t) = \left(1 - \frac{\eta \xi_C^2(t)}{\kappa^2(t-1)^2} + \frac{\sigma \xi_C^4(t)}{\kappa^2(t-1)^4}\right)\frac{\xi_C(t)}{t-1},$$

$$m_P(t) = \left(1 - \frac{\eta \xi_P^2(t)}{\kappa^2(t-1)^2} + \frac{\sigma \xi_P^4(t)}{\kappa^2(t-1)^4}\right)\frac{\xi_P(t)}{t-1}. \quad (33)$$

The last pair of relations is obtained by replacing $\xi/(t-1)$ by the almost equal $\xi_C/(t-1)$ and $\xi_P/(t-1)$ squares and higher power terms inside the brackets.

Besides, $\theta_C \approx (m_C+\xi_C)/t\kappa \approx \xi_C/\kappa(t-1)$ and $\theta_P \approx (m_P+\xi_P)/t\kappa \approx \xi_P/\kappa(t-1)$ so that it is possible to write

$$m_C(t) = \mathscr{B}_S(\theta_C),$$
$$m_P(t) = \mathscr{B}_S(\theta_P). \tag{34}$$

Hence $m_C(t)$ can be obtained using

$$\mathbf{H}_{FM}(T) = \mathbf{H}_{ex}(T)|_{\overline{J}=\overline{J}(T_C)} \doteq -(\mu B_C N + zNS^2 \overline{J}(T_C)) + \left(\frac{\overline{\varepsilon}_1(T_C)}{2} + \mu B_C\right)\frac{\hat{n}_T}{S} \tag{35}$$

where $B_C = B_{app} + m_C(T)\tau_C/\kappa\mu$, and

$$\Psi_{FM}(T) = \Psi_g(T)|_{x(T)=x_C(T)} = \prod_k |\psi_{Ck}(T)\rangle,$$

$$|\psi_{Ck}(T)\rangle = \left(\frac{1-x_C}{1-x_C^{2S+1}}\right)^{1/2} \sum_{n_k=0}^{2S} x_C^{n_k/2} |n_k\rangle \tag{36}$$

with $x_C = e^{-\theta_C(T)/S}$. The internal energy and total energy of ferromagnetism are

$$\in_{FM}(T) \doteq -(\mu B_C N + zNS^2 \overline{J}(T_C)) + N\left(\frac{\overline{\varepsilon}_1(T_C)}{2} + \mu B_C\right)[1-\mathscr{B}_S(\theta_C)],$$
$$E_{FM}(T) \doteq E_{crystal}(T) + \in_{FM}(T) \tag{37}$$

per unit volume. Similarly $m_P(t)$ can be obtained using

$$\mathbf{H}_{PM}(t) = \mathbf{H}_{ex}(t)|_{\{J_\varepsilon=0\}} = -g\mu_B B_P \sum_{j=1}^{N} S_{jz} = -\mu B_P(N - \frac{\hat{n}_T}{S}) \tag{38}$$

where $B_P = B_{app} + m_P(T)\tau_C/\kappa\mu$, and

$$\Psi_{PM}(T) = \Psi_g(T)|_{\{J_\varepsilon=0\}} = \prod_k |\psi_{Pk}(T)\rangle,$$

$$|\psi_{Pk}(T)\rangle = \left(\frac{1-x_P}{1-x_P^{2S+1}}\right)^{1/2} \sum_{n_k=0}^{2S} x_P^{n_k/2} |n_k\rangle \tag{39}$$

where $x_P = e^{-\theta_P(T)/S}$. The internal energy and total energy per unit volume of paramagnetism are

$$\in_{PM}(T) = -\mu B_P N \mathscr{B}_S(\theta_P),$$
$$E_{PM}(T) \doteq E_{crystal}(T) + \in_{PM}(T). \tag{40}$$

### D. Energy analysis for $T > T_C$

The wave function $\Psi_g(T)$ certainly differs from a simple linear combination of $\Psi_{FM}(T)$ and $\Psi_{PM}(T)$. The compounded number states $|\psi_{Ck}(T)\rangle$ and $|\psi_{Pk}(T)\rangle$ for the single mode $k$ are not orthogonal to each other. Consequently the net number states for all modes $\Psi_{FM}(T)$ and $\Psi_{PM}(T)$ are non-orthogonal. Nevertheless, a rapport can be established from expressions (30), (36) and (39) for these wave functions and the relation

$$x(t) = x_C(t)^{\mathscr{C}(t)} x_P(t)^{1-\mathscr{C}(t)} \tag{41}$$

that can be obtained from combining equations (20) for $\theta$ with (23) for $\xi(T)$ and (32) for $m(T)$: $\theta = \mathscr{C}(T)\theta_C + [1-\mathscr{C}(T)]\theta_P$. Relation (41) reveals a smooth transition of the thermal ground state from the purely ferromagnetic region (at $t \to 1-$ limit) to the purely paramagnetic regime (at $t$

→ ∞ limit). The magnetic energy from $t = 1$ to $t = \infty$ is obtained by inserting expressions (37) for $E_{FM}$ and (40) for $E_{PM}$ into equation (31):

$$E_g(T) = \mathcal{P}(t)E_{FM}(T) + [1-\mathcal{P}(t)]E_{PM}(T)$$
$$+ N\mathcal{P}(t)[1-\mathcal{P}(t)][m_C(T) - m_P(T)]\left[\frac{\bar{\varepsilon}_1(T_C)}{2} + \frac{N\tau_C}{\kappa}[m_C(T) - m_P(T)]\right]. \quad (42)$$

The last term is usually extremely small as the maximum value of $\mathcal{P}(1-\mathcal{P})$ is ¼ and $(m_C-m_P)$ is normally a very small quantity for $t > 1$, albeit being field dependent. If it is neglected, one gets a nearly accurate expression for coefficient $\mathcal{P}(t)$, namely,

$$\mathcal{P}(t) \doteq \frac{\in_{Magnon}(T) - \in_{PM}(T)}{\in_{FM}(T) - \in_{PM}(T)}. \quad (43)$$

However, $\mathcal{A}(t)$ and $[1-\mathcal{A}(t)]$ have been defined here as the linear combination coefficients of magnetization ratios $m_C(T)$ and $m_P(T)$ in equation (32). Their square roots are certainly not the coefficients of linear combination of the non-orthogonal FM and PM wave functions in the thermally equilibrated state.

## 5. Magnon-Phonon Coupling

### A. Force constants

The Born-Oppenheimer energy surface $E_g(\{R_j\}|T)$ calculated as a function of temperature accounts for lattice expansion as $T$ increases. The force on an individual center in the thermal ground state is

$$\mathbf{F}^{tot}_{j\varepsilon}(T) = -\nabla_{j\varepsilon}\left[E_{crystal}(T) + \in_{Magnon}(T)\right]$$
$$= \mathbf{F}^{crystal}_{j\varepsilon}(T) + \mathbf{F}^{magnon}_{j\varepsilon}(T) \quad (44)$$

where

$$\mathbf{F}^{crystal}_{j\varepsilon}(T) = -\nabla_{j\varepsilon}E_{crystal}(T), \quad (45)$$

and from equation (42),

$$\mathbf{F}^{magnon}_{j\varepsilon}(T) = -\nabla_{j\varepsilon}\left[\mathcal{P}(t)\in_{FM}(T) + [1-\mathcal{P}(t)]\in_{PM}(T)\right] \quad (46)$$

for $T > T_C$. The force constant is the curvature of the energy surface at the equilibrium geometry such that

$$C^\varepsilon_g(T) = \left.\frac{\partial^2 E_g(T)}{\partial R^2_{j\varepsilon}}\right|_{eq}, \quad C^\varepsilon_{crystal}(T) = \left.\frac{\partial^2 E_{crystal}(T)}{\partial R^2_{j\varepsilon}}\right|_{eq}, \quad C^\varepsilon_{mp}(T) = \left.\frac{\partial^2 \in_{SSA}(T)}{\partial R^2_{j\varepsilon}}\right|_{eq}, \quad (47)$$

all three being independent of the location ($R_j$) of the spin center in the absence of any crystal defect. It is obvious that $C^\varepsilon_g(T) = C^\varepsilon_{crystal}(T) + C^\varepsilon_{mp}(T)$. The quantity $C^\varepsilon_g(T)$ can be obtained from the computed energy values. A detailed calculation starting from equation (31) yields

$$C^\varepsilon_{mp}(T) = 2S\left[-S + \tilde{\Gamma}\left(1 - \frac{2t}{t-1}m(T)\right)\right]\ddot{J}_\varepsilon(T) \quad (48)$$

where constant $\tilde{\Gamma}$ is defined by (22) and $\ddot{J}_\varepsilon(T) = \left.\dfrac{\partial^2 J_{j\varepsilon}(T)}{\partial R_{j\varepsilon}^2}\right|_{SSA}$. Hence the magnon-phonon "force constant" can be computed. The curvature can be negative when $\ddot{J}_\varepsilon(T) > 0$ yet $2tm/(t-1) + S/\tilde{\Gamma} > 1$ that occurs as $t \to 1+$, or for $t \to 1-$, $S < \tilde{\Gamma}$ yet $\ddot{J}_\varepsilon(T) < 0$. The base line curvature $C_{crystal}^\varepsilon(T)$ can be obtained from the difference of the two computed quantities.

### B. Phonon Hamiltonian

The phonon momentum in direction $\lambda$ and the corresponding dispersion for phonon frequency are

$$q_\lambda = \frac{(-N_\lambda + 2s)}{N_\lambda \delta_\lambda} \pi, \qquad (s = 1, 2, ..., N_\lambda)$$

$$\omega_{q\lambda}^\varepsilon = \left(\frac{4C_{crystal}^\varepsilon}{\mathcal{M}}\right)^{1/2} \left|\sin \tfrac{1}{2} q_\lambda \delta_\lambda\right| \tag{49}$$

where $\mathcal{M}$ is the mass of each center. In the long wavelength limit, $\omega_{q\lambda}^\varepsilon = (C^\varepsilon / \mathcal{M})^{1/2} q_\lambda \delta_\lambda$. The displacement vectors $\delta R_{j\varepsilon}$ are in general related to the phonon eigenmodes $\delta R_{q,\lambda}$ by the polarized Fourier transformation

$$\delta R_{j\varepsilon} = N^{-1/2} \sum_{\mathbf{q},\lambda} e_{q\lambda}^\varepsilon \, \delta R_{q\lambda} \, e^{i\mathbf{q}\cdot\mathbf{R}_j}. \tag{50}$$

The term $e_{q\lambda}^\varepsilon$ with $\lambda = 1, 2, 3$ represents polarization tensor for the elastic continuum.[11] The phonon Hamiltonian reads

$$\mathbf{H}_{phonon} = \sum_{q,\lambda} \frac{\delta P_{-q,\lambda} \delta P_{q,\lambda}}{2\mathcal{M}} + \sum_{\mathbf{q},\lambda} \frac{\mathcal{M}\omega_{q,\lambda}^2}{2} \delta R_{-q,\lambda} \delta R_{q,\lambda}$$
$$= \sum_{q,\lambda} \hbar \omega_{q,\lambda}^\varepsilon \left(b_{q,\lambda}^{\varepsilon\dagger} b_{q,\lambda}^\varepsilon + \tfrac{1}{2}\right) \tag{51}$$

where $\delta P_{\mathbf{q},\lambda}$ is canonically conjugated to $\delta R_{\mathbf{q},\lambda}$ as $[\delta R_{\mathbf{q},\lambda}, \delta P_{\mathbf{q}',\lambda'}] = i\hbar \delta_{\mathbf{q},-\mathbf{q}'}\delta_{\lambda,\lambda'}$. The phonon annihilation and creation operators are given by

$$b_{q,\lambda}^\varepsilon = \left(\frac{\mathcal{M}\omega_{q\lambda}^\varepsilon}{2\hbar}\right)^{1/2} \delta R_{q\lambda} - i \frac{1}{(2\hbar \mathcal{M}\omega_{q\lambda}^\varepsilon)^{1/2}} \delta P_{q\lambda},$$

$$b_{-q\lambda}^{\varepsilon\dagger} = \left(\frac{\mathcal{M}\omega_{q\lambda}^\varepsilon}{2\hbar}\right)^{1/2} \delta R_{q\lambda} + i \frac{1}{(2\hbar \mathcal{M}\omega_{q\lambda}^\varepsilon)^{1/2}} \delta P_{q\lambda}. \tag{52}$$

The phonon energy arising from lattice vibrations is in addition to total energy $E_g$, and it can be added to $E_{crystal}$.

The change of phonon spectrum with $T$ can be visualized from expression (47). The temperature dependence enters mainly because lattice expansion weakens the crystal, decreases the force constant $C^\varepsilon$ and $q_\lambda \delta_\lambda$ per unit volume.

### C. Magnon-phonon interaction

The anisotropy-mediated magnon-phonon interaction and its small effect on the phonon spectrum accompanied by magnon decay, as well as the exchange-mediated magnon-phonon coupling and its effects such as magnon field dependence, decay rate, scattering rate and transport lifetime have been discussed in detail by Streib et al.[11] The present work instead highlights two aspects, namely, (i) temperature-dependence of magnetization in all ranges of temperature, and (ii) the way to directly evaluate the effect of the exchange-mediated magnon-phonon interaction on phonon frequencies from possible numerical calculations on model solids. In such a theory, the anisotropic effect may arise from the choice of unit cell and would be inherent in the numerical result. The general magnon-phonon interaction can be distilled from the deviation of $\mathbf{H}_{ex}$ due to lattice deviation from equilibrium whereas its numerical magnitude can be calculated from $\in_{Magnon}$ at any temperature.

The phonon frequency shift due to lattice expansion from absolute zero can be obtained from equation (49),

$$\Delta\omega_{q\lambda}^{\varepsilon}(T) = \omega_{q\lambda}^{\varepsilon}(T) - \omega_{q\lambda}^{\varepsilon}(0) = \left[\left(\frac{C_{crystal}^{\varepsilon}(T)}{C_{crystal}^{\varepsilon}(0)}\right)^{1/2} - 1\right]\omega_{q\lambda}^{\varepsilon}(0). \quad (53)$$

This is a negative quantity. The contribution of magnon-phonon coupling to phonon frequency spectrum is

$$\omega_{q\lambda,mp}^{\varepsilon}(T) = \omega_{q\lambda,ATM}^{\varepsilon}(T) - \omega_{q\lambda}^{\varepsilon}(T) = \left[\left(1 + \frac{C_{mp}^{\varepsilon}(T)}{C_{crystal}^{\varepsilon}(T)}\right)^{1/2} - 1\right]\omega_{q\lambda}^{\varepsilon}(T) \quad (54)$$

where $C_{mp}^{\varepsilon}(T)$ is given by equation (48), and it can be negative for certain phonon branches at certain temperatures. The exchange energy between two spin centers is generally much weaker than their binding energies, and the respective curvatures normally follow the same trend. For $|C_{mp}^{\varepsilon}| \ll C_{crystal}^{\varepsilon}$,

$$\omega_{q\lambda,mp}^{\varepsilon}(T) = \frac{C_{mp}^{\varepsilon}(T)}{2C_{crystal}^{\varepsilon}(T)}\omega_{q\lambda}^{\varepsilon}(T). \quad (55)$$

This contribution may be negative sometimes as discussed earlier in subsection A. These explain the observed trends in phonon frequencies and their modifications by magnon-phonon interactions.[12]

### 6. Discussion

Körmann et al.[12] estimated that the magnon-phonon interaction for some phonon modes can be an order of magnitude larger than the computed shift from lattice expansion, and concluded that the effect of magnetic short-range order persists significantly above the Curie temperature. This conclusion is in full agreement with the concept of ATM. Also, the space-spin averaging (SSA) in density functional theory (DFT) with considerations of statistical thermodynamical weights can be viewed as a methodology for computing the ATM energy. It uses the site magnon picture by default.

The work in ref. 12 was based on writing the finite temperature wave function in thermal equilibrium as a linear superposition of two extreme wave functions, $\Psi_{FM}$ ($T$=0) and $\Psi_{PM}$($T$=∞): $\Psi_{SSA}(T) \approx c_{FM}(T)\Psi_{FM}(0) + c_{PM}(T)\Psi_{PM}(\infty)$. This was obviously a drastic approximation.[25] Also, the field effect was kept obscure. The SSA energy was written as

$E_{SSA}(T) = A(T)E_{FM}(0) + [1–A(T)]E_{PM}(\infty)$ (56)

where $A(T) = c_{FM}(T)^2$ such that $A(0) = 1$, $A(\infty) = 0$, and

$$A(T) = \left[ E_{SSA}(T) - E_{PM}(\infty) \right] / \left[ E_{FM}(0) - E_{PM}(\infty) \right].$$ (57)

This simplification was possible as it was a two-state model with basis functions orthogonal to each other. It was used to formulate forces and simulate the net phonon spectrum. The formulation had two advantages: (a) The forces in the FM and PM limit are defined by the gradients of the respective energy surfaces $E_{FM}(0)$ and $E_{PM}(\infty)$ as the spin configuration in each limit is unique. (b) Temperature dependence of the adiabatic coupling parameter $A(T)$ can be obtained from quantum Monte Carlo simulations. The phonon frequency was estimated in the quasi-harmonic approximation. The effect of magnon-phonon coupling was estimated as difference from the net frequency.

The present work on a mono-domain magnet that has been based on using the correct spin-wave Hamiltonian along with only physical microstates with appropriate statistical weights possesses several distinctive features as detailed below.

- To check whether the all-temperature magnon theory indeed explains the magnetic characteristics in different temperature limits, exemplary parameters $\beta$ and $\gamma$ have been calculated for model ferromagnetic solids in Appendix 1, and representative numbers in agreement with experiment have been obtained (without higher-order magnon-magnon interactions). The critical exponents have been found to indeed vary from one ferromagnet to another and in better agreement with experiment than their mean-field theory and critical phenomenon theory counterparts. This illustrates the predictive ability of the ATM theory.
- The Weiss temperature $T_W$ has been estimated in the same Appendix, and the ratio $T_W/T_C$ is observed to rely on crystal type and the average coupling constant at $T_C$. The present work accounts for the FM magnon decay with temperature through the decay of coupling constants because of increasing lattice disorder.
- For $T > T_C$, correlation has been established between the ATM picture and FM and PM pictures, magnetization has been interpreted as a linear combination of FM and PM magnetizations, and $E_g(T)$ has been analyzed in terms of $E_{FM}(T)$, $E_{PM}(T)$ and $m(T)$ in equation (42) from which a more or less accurate expression (43) has been derived for the magnetization coefficient $\mathcal{A}(t)$. Then the force on individual spin centers in thermal equilibrium has been fully formulated and the corresponding force constant has been obtained as the sum of force constant for the "spin-less" solid and the force constant due to the Heisenberg Hamiltonian.
- It may be noticed that the grossly approximate relations (56) for energy and (57) for the square of the wave function coefficient are somewhat similar in form to (42) for total energy in equilibrium and (43) for $\mathcal{A}(T)$, the latter being a coefficient of magnetization and not directly related to the squares of wave function coefficients.

- The base line force constant gives a quantitative expression for the phonon frequency $\omega_{q\lambda}^{\varepsilon}$ that is in turn used to write down the phonon Hamiltonian. The change in phonon frequency by magnon-phonon interaction has been formulated. The frequency modification by magnon-phonon interaction can be negative for certain phonon branches near the ferromagnetic transition temperature.

This discussion completes the aims and achievements of the present work. A detailed theoretical study of the effects of anisotropy and magnon-phonon interaction as done using the traditional approach[11] is beyond the scope of the present work that is based on the correct magnon Hamiltonian and physically acceptable states, and focused on evaluation of temperature dependences and magnon-phonon coupling. A detailed numerical investigation employing DFT with a sizable number of spin configurations at varying crystal geometries is still not possible with the present-day computing abilities, [12] and remains outside the ambit of this work.

**Supplementary Information:** Supplementary Information contains Table S1 for the parameter $\Gamma$ corresponding to different types of lattice, and Table S2 showing Brillouin function constants for various spins.

**Acknowledgment**

The author gratefully acknowledges Arun K. Pal for preparing the three Figures in Appendix 1.

**Appendix 1. Trends of Basic Magnetic Properties**

Singularities are manifested by properties of a ferromagnet near the transition point.[26] For example, spontaneous magnetization follows the linear relationship $1/m \propto (1-t)^{-\beta}$ below Curie point, and susceptibility approximately satisfies $\chi = a(t-1)^{-\gamma}$ just above the transition temperature. The mean-field values of the critical exponents $\beta$ and $\gamma$ are ½ and 1, respectively, whereas theory of critical phenomena gives the improved values $\beta = 0.3645 \pm 0.0025$ and $\gamma = 1.386 \pm 0.004$.[27] In reality the observed exponents can be quite different and they vary from one ferromagnet to another. Considerably better theoretical values are realized from the ATM treatment. This is discussed next.

**1.A. Magnetization below $T_C$**

From equation (25), as $T \to T_C -$ in absence of an applied field, $\bar{\varepsilon}_1 \to 0$ and $\theta = m/\kappa t$ so that from (23) one gets

$$\theta^2 = \frac{1}{\eta}(1-t)\left[1+\frac{\sigma}{\eta^2}(1-t)+...\right] \qquad (1.1)$$

which gives the molecular field solutions. In case the second and higher order terms in right side of (1.1) are neglected, $m \approx \kappa(1 - t)^{1/2}/\eta^{1/2}$ such that $\beta = \frac{1}{2}$. This is of qualitative significance only. The general temperature dependence of $\theta$ and $m$ can be written as

$$\theta \approx (1 - t)^{\beta} \Rightarrow m \approx \kappa(1 - t)^{\beta}, \tag{1.2}$$

such that $\beta$ depends on temperature as

$$\beta = \frac{1}{2}\left[1 - \frac{\log_e \eta}{\log_e(1-t)} + \frac{\sigma}{\eta^2}\frac{(1-t)}{\log_e(1-t)}\right] < \frac{1}{2}. \tag{1.3}$$

For $t = 0.999$ one finds $\beta = 0.42$ for $S = \frac{1}{2}$, $\beta = 0.37$ for $S = 1$, and $\beta = 0.35$ for $S = 3/2$. Books on solid state physics quote typical experimental values $0.42\pm0.07$ for nickel (with $S = \frac{1}{2}$) and $0.34\pm0.04$ for iron (with $S$ varying between 1 and 3/2).

**I.B. Magnetization above $T_C$**

It is natural to expect a first order decay of coupling constants with temperature as the lattice expands: $\bar{J}(T) = e^{-\varsigma(t-1)}\bar{J}(T_C)$. This gives $\bar{\varepsilon}_1(T) = e^{-\varsigma(t-1)}\bar{\varepsilon}_1(T_C)$ as $T$ increases from $T_C$, unless there is a structural phase change when $\bar{J}(T)$ and $\bar{\varepsilon}_1(T)$ can change in a drastic way. The quantity $(\xi_C-\xi_P)$ will have the same exponential decay so that

$$\xi(T) = \xi_C - (\xi_C - \xi_P)\left[1 - e^{-\varsigma(t-1)}\right]. \tag{1.4}$$

This expression has been used here to solve equation (28) and thereafter track the variation of the inverse of susceptibility ($\chi$) with temperature. From equation (28), one finds the reduced magnetization at T=$T_C$+ as

$$m_C = \left(\frac{\kappa^2 \xi_C}{\eta}\right)^{1/3} - \xi_C. \tag{1.5}$$

The non-vanishing value of $m_C$ is an outcome of the finite applied field. Since $\kappa/\eta^{1/2} > 1$ always and normally $\xi_C < 1$, $m_C > 0$. As temperature increases $m(T)$ diminishes monotonically. Also, $m_C < 1$ unless the applied field is extremely large. Therefore, saturation magnetization is not always restored when an external field is applied at $T_C$.

Two examples are given in Table 1.1.
(1) The parameter $x_{PM} = 0.96$ (~$T_C/T_W$) is found for the spin-$\frac{1}{2}$ system that strongly resembles nickel. This gives $\bar{\varepsilon}_1(T_C) = 0.05643 k_B K$ at $B_{app} = 10$ kG. The same $x_{PM}$ is assumed for the $S = 1$ model system with the same magnetic field so that $\bar{\varepsilon}_1(T_C) = 0.1126 k_B K$. In these cases, $t_P = 1.0417$ is calculated for a fixed $\xi$. When $\xi$ is allowed to vary, $t_P$ is theoretically determined as $t_P = 1.0435$.
(2) At $B_{app} = 20$ kG one finds $\xi_P = 0.002142$ (for $S = \frac{1}{2}$) and $0.001717$ (for $S = 1$), and the corresponding $\xi_C$ values in Table 1.1. One finds $x_{PM} = 0.9794$ and $0.9795$ for the two materials. These give $t_P = 1.0209$ while $\xi$ is fixed and $t_P = 1.0214$ when $\xi$ varies.

The susceptibility $\chi$ is given by the ratio $\chi = N\mu m/B_{app}$. Since $m_C$ is finite, one writes $1/\chi = 1/\chi_C + (T - T_C)^{\gamma}/a$. This yields the well-known expression $\chi = a(T - T_C)^{-\gamma}$ when $a/N\mu$ $\ll 1$. Since $m_C > 0$ in the presence of an external magnetic field, $1/\chi$ does not blow up as $T$ approaches $T_C$ from above.

**Table 1.1** Calculated physical quantities for two spin systems at two different external magnetic fields and at temperatures greater than the ferromagnetic transition temperature.

| S | $T_C$ in K | $T_W$ in K | $\bar{\varepsilon}_1(T=T_C)$ in $k_B K$ | $B_{app}$ in $kG$ | $\xi_P$ Eq. (29) | $\xi_C$ Eq. (29) | Reduced magnetization at $T_C$ ($m_C = M(T_C)/N\mu$) |
|---|---|---|---|---|---|---|---|
| ½ | 627[a] | 654[b] | 0.0564 | 10 | 0.001071 | 0.001116 | 0.1485 |
|   |      |      |        | 20 | 0.002142 | 0.002187 | 0.1850 |
| 1 | 1043[c] | 1086[d] | 0.1120 | 10 | 0.0008584 | 0.0008942 | 0.1327 |
|   |      |      |        | 20 | 0.001717 | 0.001753 | 0.1655 |

[a] Observed data for nickel.
[b] Estimated for nickel.
[c] Observed data for iron.
[d] Estimated for iron. Average iron spin varies between 1 and 3/2.

The decay constant for $\bar{\varepsilon}_1(T)$ has been chosen as $\zeta = x_{PM}/(1- x_{PM})$ such that $\bar{\varepsilon}_1(T_W)/\bar{\varepsilon}_1(T_C)=1/e$, and subsequent numerical results are subject to this approximation. Variation of the calculated power $\gamma$ is illustrated in Figure 1.1. For most of the temperature range from $T_C$ to $T_W$, $\gamma$ varies between 1.2 and 1.4. About the same power is found for both $S$ values, and a shift to the right is observed as the field strength increases. The average power $\gamma$ for the first five points (from $t=1.005$ to $t=1.025$) is 1.345 ($S=$ ½) and 1.341 ($S=$ 1) for $B_{app} =$ 10 $kG$, and 1.326 ($S=$ ½) and 1.335 ($S=$ 1) for $B_{app} =$ 20 $kG$. To compare, the experimental values are around 1.35 for nickel and 1.33 for iron.

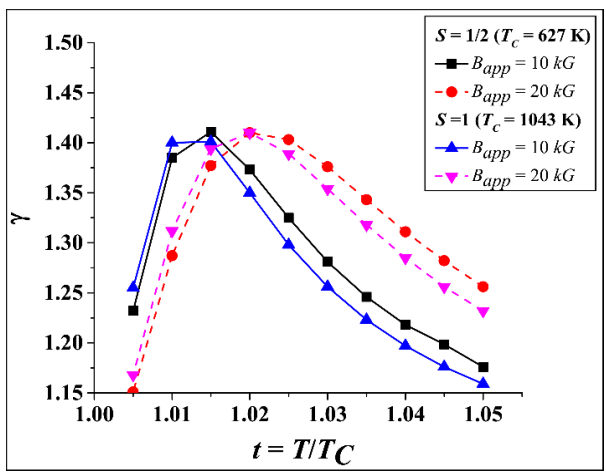

**Figure 1.1.** The calculated power $\gamma$ for variation of susceptibility as $T \to T_C$ from above. The relative temperature $t$ is defined by $t = T/T_C$, and $\bar{\varepsilon}_1(T_C)$ is taken to be independent of $B_{app}$.

The corresponding temperature dependence of the proportionality constant $a$ is shown in Figure 1.2. Unlike the power $\gamma$, the proportionality constant $a$ strongly varies with temperature, strongly depends on $S$, and goes to increase with the applied field. The average $a$ in unit $N\mu \times 10^{-4}$ is 2.52 ($S= \frac{1}{2}$) and 3.83 ($S= 1$) for $B_{app} = 10$ kG, and 2.90 ($S= \frac{1}{2}$) and 4.51 ($S= 1$) for $B_{app} = 20$ kG.

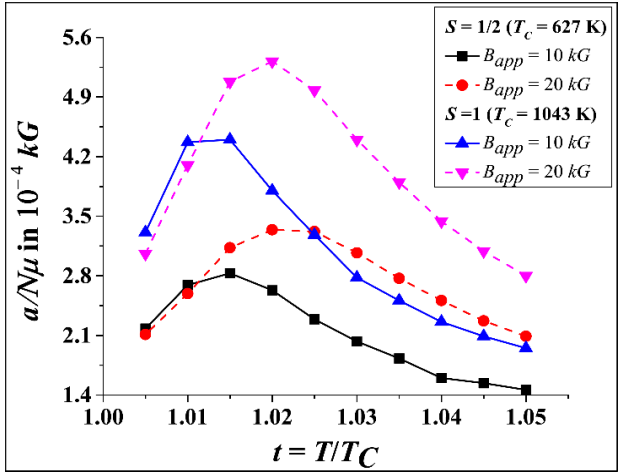

**Figure 1.2**. The calculated $a$ for variation of susceptibility as $T \rightarrow T_C$ from above. The relative temperature $t$ is defined by $t = T/T_C$, and $\bar{\varepsilon}_1(T_C)$ is taken as independent of $B_{app}$.

Of course these findings depend on the choice of the decay constant $\zeta$. Another choice would lead to somewhat different values of $\gamma$ and $a$, but the variations with respect to $T$, $S$ and $B_{app}$ must exist for finite $B_{app}$, $\bar{\varepsilon}_1(T)$ and $m_C$.

Finally, $1/\chi$ as function of $T$ is plotted in Figure 1.3. The field dependence is hardly visible as $\bar{\varepsilon}_1(T_C)$ has been assumed to be linearly proportional to $B_{app}$. The susceptibility inverse rises with the $S$ value. The asymptotes are drawn at extremely large $t$ ($t = 2.0$ with $N\mu/\chi = 9.334 \times 10^3$ kG for $S = \frac{1}{2}$ and $N\mu/\chi = 11.649 \times 10^3$ kG for $S = 1$). In both cases, $t_P = T_W/T_C$ turns out to be 1.0435 in agreement with $x_{PM} = 0.96$. If instead $\bar{\varepsilon}_1(T_C)$ was kept fixed at the values in Table A1.1, $t_P$ would have been 1.0214 for $x_{PM} = 0.98$ when $B_{app} = 20$ kG (not shown in the figure).

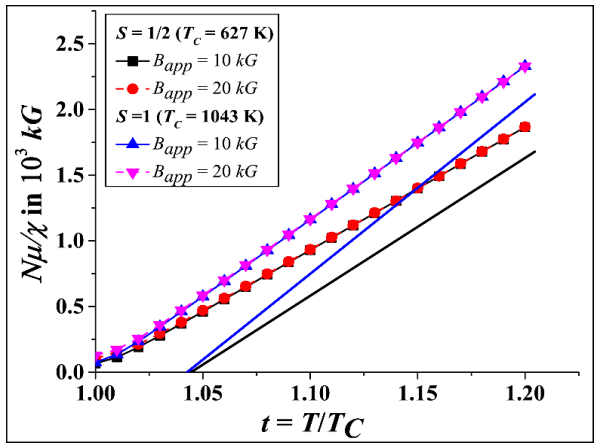

**Figure 1.3.** Inverse of susceptibility as function of temperature while $\bar{\varepsilon}_1(T_C)$ is proportional to $B_{app}$ for each $S$.

cannot be defined as a linear superposition of $\Psi_{FM}(0)$ and $\Psi_{PM}(\infty)$ at any arbitrary temperature.

**Supplementary Information**

**Table S1.** Lattice types and the parameter $\Gamma$ [a]

| Lattice Type | Number of nearest neighbors ($z$) | Examples | Parameter $\Gamma$ |
|---|---|---|---|
| I | 6 | Cubic P, Tetragonal P, Orthorhombic P, Triclinic P, Monoclinic P with β~90º, Trigonal R with α = β = γ ~ 90◦, Orthorhombic C, Monoclinic C with β ~ 90◦ | 1 |
| II | 8 | Cubic I, Tetragonal I, Orthorhombic I | 3/2 |
| III | 12 | Cubic F, Orthorhombic F | 5/3 |
| IV | 12 | Hexagonal P, Trigonal R with α = β = γ ~ 120◦ | $4/3^{2/3}$ |

[a] From ref. 14

**Table S2.** Spins and Brillouin function constants

| $S$ | ½ | 1 | 3/2 | 2 |
|---|---|---|---|---|
| $\kappa$ | 1 | 2/3 | 5/9 | ½ |
| $\eta$ | 1/3 | 1/6 | 17/135 | 13/120 |
| $\sigma$ | 2/15 | 13/360 | 26/1215 | 31/1920 |